\documentclass[3p,twocolumn,preprint, sort&compress]{elsarticle}

\journal{Physics Letters B}

\usepackage{amsmath,amsfonts,amssymb}  
\usepackage{booktabs}
\usepackage{graphicx} 
\usepackage[colorlinks=true,urlcolor=black,citecolor=blue,linkcolor=black] {hyperref}

\newcommand{\ie}{{\it i.e.}}

\newcommand{\be}{\begin{equation}}
\newcommand{\ee}{\end{equation}}
\newcommand{\br}{\begin{eqnarray}}
\newcommand{\bea}{\begin{eqnarray}}
\newcommand{\eea}{\end{eqnarray}}
\newcommand{\er}{\end{eqnarray}}
\newcommand{\ba}{\begin{array}}
\newcommand{\ea}{\end{array}}
\newcommand{\bi}{\begin{itemize}}
\newcommand{\ei}{\end{itemize}}
\newcommand{\bn}{\begin{enumerate}}
\newcommand{\en}{\end{enumerate}}
\newcommand{\bc}{\begin{center}}
\newcommand{\ec}{\end{center}}

\newcommand{\Eq}[1]{Eq.~(\ref{#1})}

\newcommand{\SU}{{\rm SU}}
\newcommand{\SO}{{\rm SO}}
\newcommand{\Tr}{\,{\rm Tr}}

\newcommand{\abs}[1]{\left |#1 \right|}

\newcommand{\Str}{\operatorname{Str}}

\newcommand{\adj}{\operatorname{adj}}
\newcommand{\pivot}{\mathcal{M}}

\renewcommand{\vec}{\mathbf}
\newcommand{\mat}{\boldsymbol}

\newcommand{\beq}{\begin{equation}}
\newcommand{\eeq}{\end{equation}}

\begin{document}

\begin{frontmatter}

\title{Light Higgs boson from multi-phase criticality in dynamical symmetry breaking}

\author[a]{Kristjan Kannike\corref{mycorrespondingauthor}}
\cortext[mycorrespondingauthor]{Corresponding author}
\ead{kannike@cern.ch}

\author[a]{Luca Marzola}

\author[a]{Martti Raidal}

\author[b]{Alessandro Strumia}

\address[a]{NICPB, R\"avala 10, 10143 Tallinn, Estonia.}

\address[b]{Universit\`a di Pisa, Dipartimento di Fisica, Italia}

\begin{abstract}
The Coleman-Weinberg mechanism can realise different phases of dynamical symmetry breaking. In each phase a combination of scalars, corresponding to the pseudo-Goldstone boson of scale invariance, has a loop-suppressed mass. We show that additional scalars, beyond the pseudo-Goldstone bosons, can become light at critical points in the parameter space where two different phases co-exist.
We present a minimal implementation of the mechanism in multi-scalar models, detailing how loop-suppressed masses and mixings can be computed.
We discuss realisations of the resulting multi-phase criticality principle and its relevance to the case of the Higgs boson.
\end{abstract}


\end{frontmatter}

\section{Introduction}

Experimental results from the Large Hadron Collider (LHC) indicate that the Higgs boson~\cite{Chatrchyan:2012ufa,Aad:2012tfa} is not accompanied by new physics responsible for stabilising the weak scale against effects arising from higher scales present in Nature, such as the Planck scale. This renews the interest in alternative ideas of naturalness and approaches aimed at understanding the origin of the electroweak scale.

One of the ideas is combining classical scale invariance, \ie{}, the absence of explicit mass terms in the scalar potential, 
with dynamical symmetry breaking. 
Gildener and Weinberg~\cite{Gildener:1976ih}, following the work by Coleman and Weinberg~\cite{Coleman:1973jx}, showed  how to approximate the dynamical symmetry breaking. Their method identifies a flat direction that arises in the scalar potential after a given combination of quartic couplings has crossed a critical condition. The quantum correction, corresponding to the renormalisation group (RG) running, then dominate the tree-level contribution along such direction and effectively shape the potential. As a result, the scalar field combination aligned with the flat direction in field space -- the dilaton pseudo-Goldstone boson of the broken classical scale invariance -- acquires a loop suppressed mass. However, no dilaton has been observed in data so far and the current experimental bounds are satisfied only in scenarios where the dilaton is heavier than the Higgs boson or mixes negligibly with it.
At the same time, alternative approaches aimed at identifying the Higgs boson with the dilaton have failed to single out the underlying mechanism.

In this work we consider dynamical symmetry breaking in a regime where additional scalars become as light as the pseudo-Goldstone bosons. This happens for special values of the parameters such that two different phases of dynamical symmetry breaking classically co-exist, and quantum effects smoothly connect them selecting the true vacuum. In the vicinity of this multi-phase critical point, the Higgs boson can be much lighter than the typical scale of new physics involved. This finding highlights the importance of critical phenomena in physics, in this particular case for explaining the smallness of  given mass scales when compared to the typical scale of symmetry breaking. 

To investigate this framework, we go beyond the Gildener-Weinberg approximation~\cite{Gildener:1976ih} and compute the loop-contributions that usually bear a negligible impact in dynamical symmetry breaking. To the contrary, when the multi-phase criticality condition is realised, these corrections are crucial to capture the physical consequences of the framework. Our computations show that extra scalar fields, not corresponding to pseudo-Goldstone bosons, acquire loop-suppressed masses while maintaining a loop-suppressed mixing with the dilaton. The effect can be understood as a consequence of a small misalignment that quantum correction induce between the particular tree-level flat direction indicated by the Gildener-Weinberg method (along which only the dilaton develops a vacuum expectation value) and the actual direction where the minimum is generated by radiative corrections (resulting in non-vanishing vacuum expectation value for all the fields).

Technically, near the multi-phase critical point, the scalar fields that develop non-vanishing vacuum expectation values, acquire masses that are independently suppressed by different $\beta$-functions, allowing for the natural emergence of a mass hierarchy. When applied to the Standard Model, the framework predicts the existence of at least a new scalar field in addition to the Higgs boson, but the mechanism allows for different mass hierarchies that can accommodate states lighter or heavier than the Higgs boson without particular tuning of the involved parameters. Because of the suppressed mixing supported by the multi-phase criticality scenario, no other signatures are expected to appear at the electroweak scale, in agreement with the LHC results.

The paper is organised as follows. In Section II we present the concept of multi-phase criticality principle (MPCP) in models of dynamical symmetry breaking. The implementation of the multi-phase criticality in minimal scenario with two scalars is presented in Section III, together with the computation of scalar masses and mixing. In Section IV we focus on the Higgs boson and present simple models that implement the proposed multi-phase criticality principle.
Finally, we conclude in Section V. Technical details related to our work are collected in appendices. Appendix A presents a matrix formalism useful to study more involved scenarios, whereas Appendix B details the computation of masses and mixing angle for the minimal model considered in the main.

\section{Multi-phase criticality in dynamical symmetry breaking}

Coleman-Weinberg symmetry breaking in theories with generic scalar quartics occurs when the RG running crosses the critical
boundary such that the tree-level scalar potential satisfies $V \ge 0$ for all field values.
Positivity implies a restriction on the quartic couplings \cite{Kannike:2012pe}, that generally is an involved intersection of multiple conditions imposed on these quantities.
Each condition corresponds to a pattern of symmetry breaking, which when realised brings the theory into some Higgsed phase. 
In each phase, a dilaton (the combination of scalars proportional to the vacuum expectation values)
has a loop-suppressed mass. 
If RG running crosses a point which is simultaneously critical for two different phases,
extra light scalars arise if the two phases are smoothly connected, rather than through a first order phase transition.

This phenomenon occurs, more in general, when different phases co-exist.
Dynamical symmetry breaking restricts which phases can be realised and how they are connected.
It is thereby interesting to study if light particles can arise at multi-phase points in this specific context.

The following concrete example illustrates that the answer is positive.
Let us consider a theory with symmetry group $G$ (for concreteness, a $\SU(N)$ group)
and one scalar $S$ in a two-index symmetric representation.
The dimensionless scalar potential, 
\beq V = \lambda ({\rm Tr}\, S S^\dagger)^2 + \lambda' \Tr S S^\dagger S S^\dagger ,
\eeq
contains two quartic couplings.
The two different critical conditions that correspond to two distinct patterns of dynamical symmetry breaking are \cite{Hansen:2017pwe,Buttazzo:2019mvl}:
\begin{itemize}
\item $\SU(N)\to \SU(N-1)$. This breaking takes place if the RG flow crosses the critical condition $\lambda + \lambda'=0$.
 
\item $\SU(N)\to\SO(N)$. This breaking takes place if the RG flow crosses the critical condition $\lambda+\lambda'/N=0$.
\end{itemize}
The two phases identified above result in different dilatons $s$ and $s'$, originally contained in $S$, which are light compared to the scale at which the breaking occurs. 
In particular, in this model the multi-critical point is trivially given by $\lambda=\lambda'=0$. Thereby, all components of $S$ become light with the exception of those eaten by possible $\SU(N)$ gauge bosons.

Less trivial situations arise in more complicated models.
For example, one scalar field in an $n$-index representation of $G$
has $n$ independent quartics that allow for $n$ inequivalent breaking patterns;
when two of them merge, one quartic can be non-vanishing,
possibly leaving massless some components of $S$ involved in the transition.
Alternatively, one can also consider models with more than one scalar representation, as we shall show next.

\begin{figure}[tb]
\begin{center}
\includegraphics[width=0.4\textwidth]{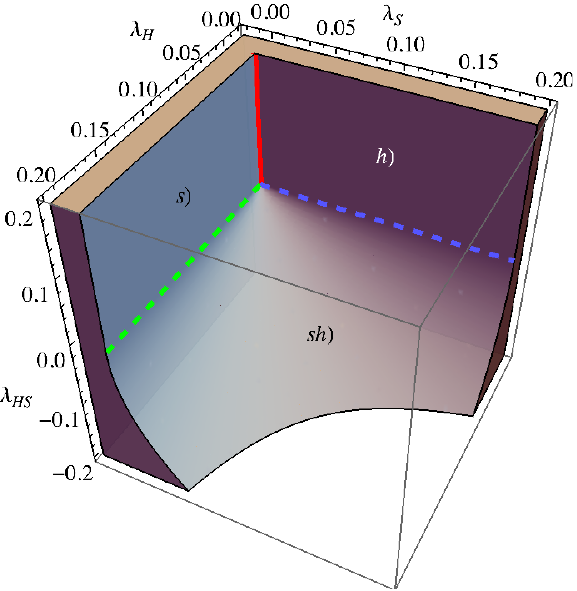}
\caption{\label{fig:CWregions}\em 
Phase structure of the model of \Eq{eq:V}.
No symmetry breaking arises in the un-shaded region, where $V\ge 0$ for all field values.
Dynamical symmetry breaking arises when RG flow of the couplings $(\lambda_H,\lambda_S,\lambda_{HS})$
crosses its boundary.
The two phases $s)$ and $sh)$ are smoothly connected along the green dashed line.
The phases $s)$ and $h)$ intersect along the red line.
}
\end{center}
\end{figure}

\section{Implementation of the multi-phase criticality principle}

\subsection{Phases in two-scalar set-up}

A light Higgs boson arises at the intersection of two phases with broken and unbroken $\SU(2)_L$. To exemplify this scenario, consider a minimal model with two scalar fields: the Higgs doublet
$H = (0,h/\sqrt{2})$ and a neutral singlet scalar $s$, with biquadratic potential
\beq 
\begin{split}
V &= \lambda_H |H|^4 +\lambda_{HS} |H|^2 \frac{s^2}{2} + \lambda_S  \frac{s^4}{4}
\\
&= \frac{1}{4} \lambda_H h^{4} +  \frac{1}{4} \lambda_{HS} h^{2} s^{2} + \frac{1}{4} \lambda_{S} s^{4}.
\end{split}
\label{eq:V}
\eeq
The couplings $ \lambda_{H}$, $\lambda_{HS} $, $ \lambda_S$ depend on the RG scale $\bar\mu$ according to the
$\beta$-functions  $\beta_X = dX/dt$ with $t = \ln(\bar\mu^2/\bar\mu_0^2)/{(4\pi)^2}$. 
We leave the $\beta$-functions generic, in order to consider more generic models with extra couplings.

The possible phases of dynamical symmetry breaking are:
\begin{enumerate}
\item[$s$)] $s\neq 0$ and $h=0$ arises when the critical boundary
\beq 
\label{eq:betacritS}
\lambda_S =0
\eeq 
is crossed, while $\lambda_{HS}>0$ 
gives a tree-level positive squared mass to the Higgs boson.
In this phase the two scalars are not mixed.
Dynamical symmetry breaking happens if $\beta_{\lambda_S}>0$ along the critical boundary.

\item[$h$)] $h\neq 0$ and $s=0$ arises when $\lambda_H =0$ and $\lambda_{HS}>0$.
In this phase the two scalars are not mixed.
This is the case originally considered by Coleman and Weinberg, now
excluded by the Higgs mass measurement that implies $\lambda_H \approx 0.12$.

\item[$sh$)] $s,h\neq 0$ arises when the critical boundary
 \beq \label{eq:shcond}
2\sqrt{\lambda_H \lambda_S}+\lambda_{HS} =0\eeq is crossed, while $\lambda_{HS}<0$, $\lambda_{H,S}\ge 0$.
The flat direction is  given by $s/h = (\lambda_H/\lambda_S)^{1/4}$.
Dynamical symmetry breaking happens if 
\beq 
\label{eq:betacrit}
\beta_{\rm crit}=\lambda_S \beta_{\lambda_H} + \lambda_H \beta_{\lambda_S} -\lambda_{HS} \beta_{\lambda_{HS}}/2 >0
\eeq 
along the critical boundary. 
In this phase the two mass eigenstates are admixtures of the original scalar fields.

\end{enumerate}
Fig.~\ref{fig:CWregions} shows the three critical boundaries geometrically.

The phases $h)$ and $s)$ are not smoothly connected.
They correspond to two different flat directions separated by the tree-level potential $V=\lambda_{HS} h^2 s^2/4 \ge 0$.
At their intersection (represented by the red line in Fig.~\ref{fig:CWregions})
the potential has two disjoint local minima with $h\neq 0$ and with $s\neq 0$, 
corresponding to a first-order phase transition with no extra light scalars.

On the other hand, Fig.~\ref{fig:CWregions} shows that the phases $s)$ and $sh)$ are smoothly connected
along the dashed green line.
Indeed, they correspond to a common flat direction, the field $s$.
The squared Higgs mass changes sign between the phases $s$) and $sh$). 
Therefore, the Higgs boson is light around their intersection.
Furthermore, as the scalar mixing vanishes in the phase $s)$, it must be small near the multi-phase point.
The two conditions in Eqs.~\eqref{eq:betacritS} and \eqref{eq:shcond} intersect at
\beq\label{eq:multiphase} \lambda_S(\bar\mu)=\lambda_{HS}(\bar\mu)=0, \eeq 
that trivially implies a massless Higgs boson.

The running of the Higgs quartic coupling in the standard model (SM) alone, 
where $\beta_{\lambda_H}|_{\rm SM} \approx - 2$ around the weak scale, 
does not cross the boundary condition for $sh)$ in the direction that gives dynamical symmetry breaking. Consequently, in models where the SM running is dominant, the Higgs can acquire a vacuum expectation value only for small values of $\lambda_S$ close to the multi-phase critical point.

\subsection{Computing scalar masses and mixing}
\label{massescomp}

Having made clear the gist of the scenario, we now compute the vacuum expectation values, masses and mixing angles of the  scalars involved in the two-field model of \Eq{eq:V}.
A more general and abstract computation is presented in \ref{app:matrix:min}.

We start by summarising the usual Coleman-Weinberg computation for the phase $sh)$, 
as it provides an example about how dynamical symmetry breaking can be approximated
using the RG-improved tree-level potential alone, omitting the more complicated and model-dependent full one-loop contribution.
Along the tree-level flat direction the potential can be approximated by replacing the tree-level potential
with field-dependent effective couplings expanded at the first order in the beta functions: 
 \be
 \lambda_{\rm eff}(s') = \lambda + \beta_\lambda \ln (s^{\prime 2}/s_0^2),
 \ee
 where $s_0$ is the typical scale of the problem and $s^{\prime2} = s^2 + h^2$ is the distance in field space.
We stress that this approximation is appropriate along the flat direction, rather than in all field space.

 The minimum approximately lies on the flat direction, at field dependent scale $s^{\prime 2}=s^2_0/e^{1/2}$ determined by the flat direction scale $s^2_0$. The eigenvalues of the field-dependent mass matrix evaluated at the minimum of the potential are
\beq 
m_{s'}^2 = \frac{2 s_0^2}{(4\pi)^2e^{1/2}} \frac{\beta_{\rm crit}}{\lambda_H+ \lambda_S  - \lambda_{HS}},
\quad m_{h'}^2 =-\frac{s_0^2\lambda_{HS}}{e^{1/2}},
\eeq
where $s'$ is the dilaton, and $h'$ is the combination orthogonal to $s'$.
Within the Coleman-Weinberg approximation, $h'$  receives a dominant tree-level mass contribution,
and the approximation does not allow to include loop-suppressed corrections.
However, at the multi-critical point under examination, the tree-level contribution to $m_{h'}^2$ is also vanishing or
small enough that the one-loop corrections must necessarily be retained.

In regard of this, the general form of the one-loop potential is
\begin{equation}
  V = V^{(0)} + V^{(1)},
\label{eq:V:eff}
\end{equation}
with the tree-level part $V^{(0)}$ given in \Eq{eq:V}, having omitted terms involving other possible scalars. 
The one-loop term, $V^{(1)}$, is given by
\beq \label{eq:V1}
V^{(1)}|_{\overline{\rm MS}} = \frac{1}{4(4\pi)^2} \Tr \bigg[M_S^4 \left( \ln \frac{M_S^2}{\bar\mu^2} -\frac32 \right)+\qquad\eeq
$$\qquad\qquad- 2 M_F^4 \left(\ln\frac{M_F^2}{\bar\mu^2} -\frac32 \right)+ 3 M_V^4 \left( \ln\frac{M_V^2}{\bar\mu^2}-\frac56 \right)\bigg].$$
Here, $\bar\mu$ indicates the RG scale introduced by the regularisation.
We used the $\overline{\rm MS}$ dimensional regularisation scheme.
The parameters in the tree-level potential also depend on $\bar\mu$ as dictated by their RGEs
(this is sometimes called RG improvement).
This cancels the dependence of the potential on the arbitrary parameter $\bar\mu$,
up to higher-loop orders and up to wave-function renormalisation. The symbols $M_{S,F,V}$ denote the usual field-dependent masses of generic scalars, fermions and vectors, respectively.
For example, $M_V^2 = g_h^2 h^2 + g_s^2 s^2$ is the mass of the U(1) gauge boson in a model where $h$ and $s$ have corresponding gauge charges $g_h$ and $g_s$. 
In more general models $M^2_{S,F,V}$ are mass matrices and their
eigenvalues (needed to compute \Eq{eq:V1}) often do not have a useful closed analytic form.
For example, complicated expressions for $M_F$ easily arise in models where $h$ and $s$ have Yukawa couplings to fermions.
Rather than resorting to model-dependent numerical methods, we next derive simple analytic expressions.

Eq.~\eqref{eq:V1} can be simplified taking into account that $h \ll s$ at the multi-phase critical point, 
so that all field-dependent masses $M$ 
acquire a common form,
\beq 
\label{eq:Mapprox}
M_i \simeq c_i s + c'_i h^2/s + \cdots.
\eeq
The above expansion fails in regions of the parameter space
where couplings have values that undo the $h \ll s$ hierarchy. In such a case the Higgs mass is still loop-suppressed but it is given by a
more complicated and model-dependent expression, therefore
we will not consider this possibility.
Expanding $V^{(1)}$ up to quartic order in $h$ shows that  the full potential
(not restricted to the flat direction) is well approximated
by the tree-level potential with the $\lambda_S$ and $\lambda_{HS}$ couplings replaced by 
\begin{equation}
\lambda_S^{\rm eff}(s) = \frac{{\beta}_{\lambda_S }}{(4\pi)^2}   \ln \frac{s^2}{s^2_S },\qquad
\lambda_{HS}^{\rm eff}(s) = \frac{{\beta}_{\lambda_{HS} }}{(4\pi)^2} \ln \frac{s^2 }{s^2_{HS}},
\label{eq:run:lambda}
\end{equation}
where $\beta_{\lambda_S}$ and ${\beta}_{\lambda_{HS} }$ are the $\beta$-functions.
The contributions to 
the $\beta$-functions coming from wave-function renormalisations of $s$ and $h$, not included in the one-loop potential, 
vanish as we are expanding around $\lambda_S=\lambda_{HS}=0$.
The $s_S$ and $s_{HS}$ parameters are computed by expanding the one-loop potential,
obtaining the usual logarithmic running plus non-logarithmic terms,
that can be traded for the precise scales $s_S$ and $s_{HS}$
(as opposed to a generic typical scale $s_0$)
in the logarithmic terms only.
They effectively encode how the full one-loop result deviates from the multi-phase critical point
hit by the RG running.
Given that the one-loop potential in the approximation of Eq.~\eqref{eq:Mapprox}
becomes similar to the running potential, the two parameters $s_S$ and $s_{HS}$
loosely correspond to the RGE scales at which $\lambda_S(\bar\mu)$ and $\lambda_{HS}(\bar\mu)$ vanish, respectively.
The precise order-one 
value of $R = e^{-1/2} s_S^2/s_{HS}^2$ encodes how much the full result deviates from a na\"{\i}ve  
running potential.

Assuming that the $\beta$-functions of $\lambda_S$ and $\lambda_{HS}$ are comparable and much smaller than $\lambda_H$,
the potential has a minimum at non-vanishing $s$ and $h$ 
\beq s \approx e^{-1/4} s_S ,\qquad
h \approx 
\frac{ e^{-1/4} s_S}{4\pi} \sqrt{\frac{-\beta_{\lambda_{HS}}\ln R}{2\lambda_H}},
\eeq
provided that $-\beta_{\lambda_{HS}}\ln R>0$
(otherwise only $s$ acquires a vacuum expectation value). 
The mass eigenvalues are both loop-suppressed:
\beq m_s^2 \approx \frac{2s^2 \beta_{\lambda_S}}{(4\pi)^2},\qquad
m_h^2 \approx \frac{-s^2 \beta_{\lambda_{HS}}\ln R}{(4\pi)^2} = 2\lambda_H h^2 .
\eeq
Their mixing angle is also loop-suppressed,
\beq
\theta \approx \sqrt{-\frac{\beta_{\lambda_{HS}}^3 \ln R}{2\lambda_H}} 
\frac{1+\ln R}{4\pi(2\beta_{\lambda_S} + \beta_{\lambda_{HS}} \ln R)},
\eeq
unless the two scalars are nearly degenerate.
No combinations of $M_h, M_s,\theta$ are univocally predicted.
All these results get multiplicatively corrected by the wave-function renormalisation effects that can be neglected.

Here we have outlined the computation in a specific model with the minimal field content.
Similar computations along these lines can be performed in more general models.
In \ref{app:matrix:min} we show how the results get extended.
In \ref{app:detail:min} we show how the general formalism can be applied to recover
the results for the specific model considered in this section.

\section{Simple models}\label{minmod}
In the minimal model with just the singlet scalar $s$ in addition to the Higgs boson $H$ and the rest of the SM,
the RG equations are given by
\begin{subequations}
\begin{align}
\label{sys:RGEmin}
 \beta_{\lambda_S} &= 9 \lambda_S^2 +\lambda_{HS}^2 , \\
\beta_{\lambda_{HS}} &= \lambda_{HS}   \left[   \tilde Z_h  + 3\lambda_S+2\lambda_{HS}\right] ,\\
 \beta_{\lambda_H} &=  \beta_{\lambda_H}^{\rm SM}+\frac14 \lambda_{HS}^2,
\label{eq:betalambdaH}
\end{align}
\end{subequations}
where the $\beta$-function
\beq
 \beta_{\lambda_H}^{\rm SM}= 2 \tilde Z_h  \lambda_H  -3 y_t^4
+\frac{9 g_2^4}{16}+\frac{27 g_1^4}{400}+\frac{9 g_2^2 g_1^2}{40}
\eeq
is the SM contribution to the running of $\lambda_H$.
Here $\tilde Z_h = Z_h + 6 \lambda_{H}, $ and
\beq
Z_h = 3 y_{t}^2- \frac{9}{4} g_2^2  -\frac 34  g_Y^2
\eeq
is the one-loop wave-function renormalisation for the Higgs boson.
As expected, it cancels out in Eq.~\eqref{eq:betacrit}.

\begin{figure}[tb]
\begin{center}
\includegraphics[width=0.42\textwidth]{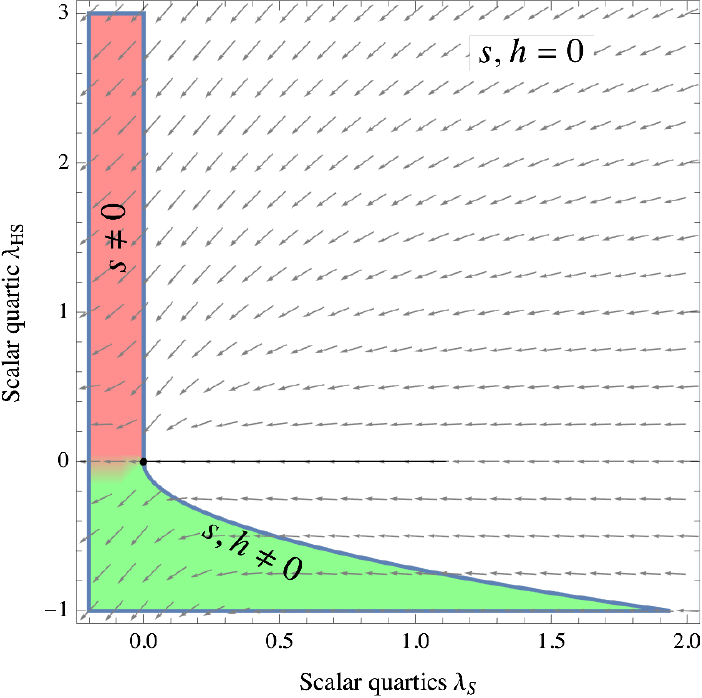}
\caption{\label{fig:RGEflowFixedPoint}\em In the minimal model,
the RG flow towards low energy 
that starts from $\lambda_S \gg \lambda_{HS}\gg 1$ is attracted towards
the multi-phase point  (dot) where $\lambda_S$ and $\lambda_{HS}$ are small.
}
\end{center}
\end{figure}

Of course, one possible reason for being near the multi-phase critical point of \Eq{eq:multiphase}
is just a fine-tuning of the parameters.
Here we discuss one possible dynamical motivation for such tuning, namely
that the RG flow converges towards $\lambda_S\simeq \lambda_{HS} \simeq 0$.
Taking into account the $\lambda_S$ and $\lambda_{HS}$ couplings only,
Fig.~\ref{fig:RGEflowFixedPoint} shows that their RGE flow towards low energy 
can bring them nearer to the multi-phase critical point $\lambda_S = \lambda_{HS}=0$.
A near-approach can arise starting from $\lambda_S \gg |\lambda_{HS}|$ at high energy, which can be justified as follows.

Assuming that $s$ arises as a massless composite scalar
corresponding to an operator ${\cal O}$ in some fundamental theory,
one can introduce a separate field $ s_0$ for the composite state by adding to the 
fundamental Lagrangian renormalised at the compositeness energy scale $M_{\rm UV}$ the term
\beq Z_0 (\partial_\mu s_0)^2/2 + (s_0-{\cal O})^2\eeq
with a vanishing kinetic term $Z_0=0$ such that $s_0$ is a Lagrange multiplier~\cite{Weinberg:1962hj}.
Interactions generate a small $Z$, such that the canonical field $s =s_0 \sqrt{Z}$ obtained from
the bare field $s_0$ has non-perturbatively large 
\beq \lambda_S(M_{\rm UV})\propto 1/Z^2\,\,\, \text{ and }\,\,\, \lambda_{HS}(M_{\rm UV})\propto 1/Z \eeq
while the quartic $\lambda_H$ of the Higgs is unconstrained, as
no composite particles are involved in $|H|^4$. Among the many composite models that one can consider, we mention the possibility that $s$ could be the
``conformal mode of the graviton'' obtained reducing $f(R)$ theories to the Einstein basis.
In theories where $f(R) =  -\frac12 \bar M_{\rm Pl}^2 R + R^2/3f_0^2$
the scalar $s$ has renormalisable quartics, $\lambda_S = f_0^2$, $\lambda_{HS} = f_0^2 (\xi_H + 1/6),$ 
where $\xi_H$ is the non-minimal coupling of the Higgs boson to gravity, and
$f_0$ becomes non-perturbative at large energy~\cite{Salvio:2017qkx}.

However, reducing $\lambda_{HS}$ down to a loop-suppressed value of order $1/(4\pi)^2$ needs
a long running of about $(4\pi)^2$ orders of magnitude (thereby, much above the Planck scale and even above
the scale where hypercharge gets strongly coupled), near to
a fixed flow of the RGE system~\cite{Giudice:2014tma} which is not infrared attractive.

\allowdisplaybreaks

In the minimal model RG evolution does not change the sign of $\lambda_{HS}$.
Crossing $\lambda_{HS}=0$ becomes possible adding extra vectors under which $H$ and $s$ are charged (but introducing extra $Z'$ vectors risks forbidding the SM Higgs Yukawa couplings), or extra fermions (but this tends to contribute as $\beta_{\lambda_{HS}}<0$) or extra scalars.
Following the latter option, we introduce a second scalar $s'$.
The most general quartic potential symmetric under $\mathbb{Z}_2\otimes\mathbb{Z}'_2$ 
(that respectively act as $s\to - s$ and as $s'\to - s'$) is
\begin{eqnarray}
  V &=& \lambda_{H} |H|^{4} + \frac{\lambda_{S} }{4} s^{4} + 
  \frac{\lambda_{S'} }{4} s^{\prime 4} +
  \\
  &&+  \lambda_{HS} |H|^2 \frac{s^{2}}{2}  +  \lambda_{HS'} |H|^2  \frac{s^{\prime 2}}{2} 
  + \frac{\lambda_{SS'} }{4} s^{2} s^{\prime 2} .\nonumber
\end{eqnarray}
In this model the $\beta$-functions are 
\begin{subequations}
\begin{align}
\label{sys:RGEmodelHSS'}
  \beta_{\lambda_H} &= \beta_{\lambda_H}^{\rm SM} + \frac{1}{4} (\lambda_{HS}^2 +  \lambda_{HS'}^2),
  \\
  \beta_{\lambda_S} &= 9 \lambda_{S}^2 +  \frac14 \lambda_{SS'}^2  +  \lambda_{HS}^2,
  \\
  \beta_{\lambda_{S'}} &= 9\lambda_{S'}^2  + \frac14 \lambda_{SS'}^2 +  \lambda_{HS'}^2,
  \\
  \beta_{\lambda_{HS}} &= \lambda_{HS} (\tilde Z_h  + 3\lambda_S+2\lambda_{HS})+\frac12 \lambda_{SS'} \lambda_{HS'},
  \\
 \beta_{\lambda_{HS'}} &= \lambda_{HS'}(\tilde Z_h  + 3\lambda_{S'}+2\lambda_{HS'})  + \frac12 \lambda_{SS'} \lambda_{HS},
  \\
 \beta_{\lambda_{SS'}} &=  \lambda_{SS'} (2 \lambda_{SS'} + 3 \lambda_{S}+
      3 \lambda_{S'}) + 2 \lambda_{HS} \lambda_{HS'}.\qquad\quad
\end{align}
\end{subequations}
For simplicity we only consider the phases where $s'=0$,
for which it is sufficient to have $\lambda_{S'},\lambda_{HS'} ,\lambda_{SS'} > 0$.
We then have a positive contribution to $\beta_{\lambda_{HS}}$, so that 
$\lambda_{HS}$ can run negative at low energy.

As a final aside comment, we mention 
another possible rationale for being near to the multi-phase critical point $\lambda_H=\lambda_{HS}=0$:
an RG flow with $\lambda_{HS}$ that runs much faster than $\lambda_S$, down to negative values at low energy.
Then the symmetry breaking critical condition in Eq.~\eqref{eq:shcond} could be crossed
only near to $\lambda_H=\lambda_{HS}=0$. 
However we do not know how to realise this flow without generating other problems.\footnote{For example, one can
reduce the symmetry of the model to a single $\mathbb{Z}_2$
(under which both $s$ and $s'$ flip sign), such that the
extra coupling $ \lambda_{HSS'}\, |H|^2 ss'$ is allowed.
If  larger than other couplings, $ \lambda_{HSS'}$ provides the desired flow.
The problem is that a dominant quartic $\lambda_{HSS'}$ modifies the dynamical symmetry breaking conditions \cite{Kannike:2016fmd}:
by itself it implies vacuum expectation values $ss'<0$.}

\section{Conclusions}
Dynamical symmetry breaking is a specific form of symmetry breaking in $3+1$ space-time dimensions
that:  leads to a pseudo-Goldstone boson (the dilaton) with loop-suppressed mass;
restricts the parameters of possible broken phases; predicts how they are connected.
We studied how the dynamical symmetry breaking  behaves
near the multi-phase critical boundaries, where different patterns of symmetry breaking may co-exist.
We found that some phases are smoothly connected, so that extra scalars become light when such phases merge.
We explored the possibility that the lightness of the Higgs boson follows from the multi-phase criticality in
dynamical symmetry breaking. 
We found that it can be realised in the example models considered in this work, 
around a specific multi-phase critical point where specific scalar couplings are  vanishingly small, $\lambda_S=\lambda_{HS}=0$.
As a result, not only the Higgs boson acquires a loop-suppressed mass similarly to the dilaton, 
but also its mixing with the dilaton is loop-suppressed, allowing to satisfy collider bounds.

We have shown how the simple Gildener-Weinberg approximation can be extended to obtain
simple expressions for the loop-suppressed masses and mixing.
The original Gildener-Weinberg flat direction in the multi-field space is misaligned from the axis by a small tilt --- an effect that is usually negligible.
In this case, however, it contributes to the mass of the Higgs boson and to its mixing with extra particles.

In this framework the Higgs boson and the dilaton masses are both loop-suppressed,
being proportional to square roots of different combinations of $\beta$-functions.
As couplings beyond the SM are unknown, this scenario gives no univocal prediction, allowing for
different mass orderings between the Higgs boson and the dilaton. 
The new physics associated with the dynamical symmetry breaking can be around the weak scale
and weakly coupled to the Higgs, in agreement with collider bounds.
The additional new scalar, the dilaton, must be discovered to test our framework.

\subsection*{Acknowledgements}
This work was supported by the ERC grant 669668 NEO-NAT, by European Regional Development Fund through the CoE program grant TK133, 
the Mobilitas Pluss grants MOBTT5, MOBTT86, and the Estonian Research Council grants PRG434, PRG803 and PRG356.

\appendix 

\section{General Matrix Formalism}
\label{app:matrix:min}
In this appendix we extend the multi-phase point computation presented in section~\ref{massescomp} 
in a specific simple model to a generic scalar field content $\phi_i$,
adapting the matrix formalism of~\cite{Kannike:2019upf,Kannike:2020ppf}. 
The effective potential is again given by \Eq{eq:V:eff},
where $V^{(0)}$ is the tree-level and $V^{(1)}$ is the one-loop contribution.
We define the scalar field vector $\vec{\Phi} = \{\phi_1, \ldots,\phi_n\}$.
We consider a tree-level quartic potential $V^{(0)}$ biquadratic in the fields, 
such that  quartic couplings form a symmetric matrix $\mat{\Lambda}$ as
\begin{equation}
  V^{(0)} = (\vec{\Phi}^{\circ 2})^{T} \! \mat{\Lambda} \vec{\Phi}^{\circ 2}.
\end{equation}
Here $\vec{\Phi}^{\circ 2}= \{\phi_1^2, \ldots,\phi_n^2\}$
is the the Hadamard square $\vec{\Phi}\circ\vec{\Phi}$ of the $\vec{\Phi}$ vector.\footnote{In general, the Hadamard product is the component-wise product of two matrices: $(\mat{A} \circ \mat{B})_{ij} = A_{ij} B_{ij}$.}  
This simplification allows to write a necessary and sufficient condition for the stability of the tree-level scalar potential,
$V^{(0)} \ge 0$ for all $\vec{\Phi}$.
The condition is that the matrix $\mathbf{\Lambda}$ be copositive~\cite{Kannike:2012pe}. 
According to the Cottle-Habetler-Lemke (CHL) theorem~\cite{Cottle1970295},  
$\mathbf{\Lambda}$ is copositive if 
\begin{itemize}
\item[1)] the principal submatrices of order $n-1$ of $\Lambda$ are copositive.
These are the matrices obtained from $\mathbf{\Lambda}$ deleting  one row and the corresponding column.
\end{itemize}
and
\begin{itemize}
\item[2)] either 
\begin{itemize}
  \item[2a)] $\det (\mathbf{\Lambda}) \geq 0$; 
  \end{itemize}
or
\begin{itemize}
  \item[2b)] at least one element of $ \adj (\mathbf{\Lambda})$
  is negative. The adjugate is related to the matrix inverse as $ \mat{\Lambda}^{-1} =  \adj (\mathbf{\Lambda})/\det(\mathbf{\Lambda})$, but exists even when $\det(\mathbf{\Lambda})=0$.
\end{itemize}
\end{itemize}
This iterative procedure implies that all diagonal elements of $\mathbf{\Lambda}$ must be $\ge 0$, and
for $n \ge 2$ adds a complicated set of extra conditions, corresponding to the various phases of the theory.

The one-loop potential can be written as
\begin{equation}
\begin{split}
    V^{(1)} &= 
 \mathbb{A} + \mathbb{B} \ln \frac{\pivot^{2}}{\bar\mu^{2}},
\end{split}
\label{eq:V:(1)}
\end{equation}
where 
\begin{align}
  \mathbb{A}  &= \frac{1}{64 \pi^{2}} \Str \mat{M}^{4} \left(\ln \frac{\mat{M}^{2}}{\pivot^{2}} - \mat{C} \right), 
\label{eq:A}
\\
 \mathbb{B} &= \frac{1}{64 \pi^{2}} \Str \mat{M}^{4}.
\label{eq:B}
\end{align}
The matrix $\mat{M}^{2}$ comprises the field-dependent matrices of scalars, vectors and fermions that appear in the theory, while $\mat{C}$ (in the $\overline{\rm MS}$ scheme) is a constant diagonal matrix with entries $3/2$ for scalars and fermions and $5/6$ for vector bosons.
In order to split the one-loop contributions, we introduced an arbitrary field-dependent ``pivot scale'' $\pivot $ with the dimension of mass \cite{Chataignier:2018aud}. 
We choose $\pivot^2 = \vec{e}_{\pivot}^T \vec{\Phi}^{\circ 2}$, where $\vec{e}_{\pivot}$ is a constant vector,
to be conveniently chosen.

The RG scale $\bar\mu$ multiplies the $\mathbb{B}$ term only, and it cancels with the running
of the couplings in the tree-level potential.
Indeed, the Callan-Szymanzik equation tells that
\begin{equation}
  \frac{d V^{(0)}}{d t} = (4 \pi)^{2} \mathbb{B} = (\vec{\Phi}^{\circ 2})^{T} \! \mat{\beta} \vec{\Phi}^{\circ 2} - \vec{\Phi}^{T} \mat{\gamma} 
  \nabla_{\vec{\Phi}} V^{(0)}\label{eq:CZ}
\end{equation}
where 
\begin{equation}
\mat{\beta} = \frac{d\mat{\Lambda}}{dt},\qquad  t = \frac{ \ln(\bar\mu^{2}/\bar\mu_{0}^{2}) }{(4 \pi)^2}
\end{equation}
is the matrix of $\beta$-functions. 
The anomalous dimension matrix $\mat{\gamma}$, which is diagonal in the biquadratic case, accounts for wave-function renormalisation.

By setting $\bar\mu = \pivot$, so that $\pivot_0 = \bar{\mu}_0$, the effective potential becomes
\begin{equation}
  V(t) = V^{(0)}(t) + \mathbb{A},
  \label{eq:Veff}
\end{equation}
where $V^{(0)}(t)$ is the tree-level potential with field-dependent effective couplings $\mat{\Lambda}(t)$, but we neglect anomalous dimensions.\footnote{In general, fields in Eq.~\eqref{eq:Veff} must be scaled with anomalous dimensions as $\vec{\Phi}(t) = \exp(\mat{\Gamma}(t)) \vec{\Phi}(t_{0})$, where  $\mat{\Gamma}(t) = -\int_{t_{0}}^{t} \mat{\gamma}(s) ds$. Then, to avoid ambiguity, one can denote $\vec{\Phi} \equiv \vec{\Phi}(t_{0})$ and take the anomalous dimensions into account by scaling $\mat{\Lambda} \to \exp(2\mat{\Gamma}(t)) \mat{\Lambda} \exp(2 \mat{\Gamma}(t))$ and $\mat{\beta} \to \exp(2\mat{\Gamma}(t)) \mat{\beta} \exp(2 \mat{\Gamma}(t))$. 
However, multi-phase conditions usually demand the vanishing of the relevant couplings multiplicatively 
corrected by wave function renormalisation.
In addition, the second term in Eq.~\eqref{eq:CZ} that depends on $\mat{\gamma}$ is proportional to $h^{2}$ and therefore small.} The running parameter is now also $\pivot$-dependent:
\begin{equation}
  t = \frac{1}{(4 \pi)^{2}} \ln \frac{\pivot^{2}}{\pivot_{0}^{2}}.
\end{equation}

The minimisation procedure can be simplified by 
considering only the submatrix in $\mat{\Lambda}$ selected by the subspace of fields which acquire non-vanishing vacuum expectation values, ignoring the others.
The stationary point equation then is
\begin{equation}
  \vec{0} = \nabla_{\vec{\Phi}} V = 4 \vec{\Phi} \circ \mat{\Lambda}(t) \vec{\Phi}^{\circ 2} 
  + \nabla_{\vec{\Phi}} \mathbb{A} + \frac{d V}{dt} \nabla_{\vec{\Phi}} t.
  \label{eq:minmf}
\end{equation}
We have $\nabla_{\vec{\Phi}} \pivot^{2} = 2 \vec{e}_{\pivot} \circ \vec{\Phi}$ and consequently
\begin{equation}
  \nabla_{\vec{\Phi}} t = \frac{2}{ (4\pi)^{2}} \frac{1}{\pivot^{2}} \vec{e}_{\pivot} \circ \vec{\Phi}.
  \label{eq:min:vector}
\end{equation}
The \emph{radial} minimisation equation is obtained projecting Eq.~\eqref{eq:minmf} along the field vector:
\begin{equation}
\begin{split}
  0  =  \vec{\Phi}^{T} \nabla_{\vec{\Phi}} V &= 4 V^{(0)}(t) + 4 \mathbb{A} + 2 \mathbb{B}
= 4 V + 2 \mathbb{B},
\end{split}
\label{eq:min:rad}
\end{equation}
where we used $\vec{\Phi}^{T} \nabla_{\vec{\Phi}} \mathbb{A} = 4 \mathbb{A}$, which holds because $\mathbb{A}$ is a homogenous function of order four. 

Let us now require that $V(t_{0}) = 0$ at a scale $t_{0}$ (where the tree-level flat direction along which $V^{(0)}=0$ is recovered) and that $\mathbb{B} > 0$, to ensure that the potential is bounded from below at higher scales. 
Then Eq.~\eqref{eq:min:rad}, expanded to linear order in $t$, takes the form
\begin{equation}
  0 \approx 4 \left(V(t_{0}) + \frac{d V^{(0)}}{dt} t\right) + 2 \mathbb{B},
\end{equation}
which recovers the familiar Gildener-Weinberg relation $\pivot^{2} = e^{-1/2} \pivot_{0}^{2}$ between the pivot scale at a stationary point and at the flat direction, \ie $t = t_{0} - 1/ [2(4 \pi)^{2}]$.

Using  Eq.~\eqref{eq:CZ}, we can  write Eq.~\eqref{eq:minmf} as
\begin{equation}
  \vec{0} = \vec{\Phi} \circ \left( 4 \mat{\Lambda}(t) \vec{\Phi}^{\circ 2} + 2 \mathbb{B} \frac{\vec{e}_{\pivot}}{\pivot^{2}}  \right),
  \label{eq:min:vector:2}
\end{equation}
where we have momentarily ignored the $\nabla_{\vec{\Phi}} \mathbb{A}$ term for simplicity. This correction, in fact, does not alter the analytical form of the solution and its importance depends on the choice made for the pivot scale which enters in $\mathbb{A}$. We will return to this point below and provide an explicit example of the treatment of these corrections for the two-fields model discussed in the main text.  

We search the stationary point at $\vec{\Phi}\neq 0$.
Using the radial Eq.~\eqref{eq:min:rad}, the factor in parentheses in Eq.~\eqref{eq:min:vector:2} gives 
\begin{equation}
  \mat{\Lambda}(t) \vec{\Phi}^{\circ 2} = \frac{V}{\pivot^{2}} \vec{e}_{\pivot},
\label{eq:vector:nonzero:vev}
\end{equation}
solved by
\begin{equation}
 \vec{\Phi}^{\circ 2} = \frac{1}{\det( \mat{\Lambda}(t) )} \frac{V}{\pivot^{2}} \adj( \mat{\Lambda}(t) ) 
 \vec{e}_{\pivot}.
 \label{eq:sol:impl}
\end{equation}
Approximating $V$ by $V^{(0)}$ this becomes
\begin{equation}
  \vec{\Phi}^{\circ 2} =  \frac{\mathcal{M}^{2}}{\vec{e}_{\mathcal{M}}^{T} \! \adj (\mat{\Lambda}(t)) \vec{e}_{\mathcal{M}}} \adj (\mat{\Lambda}(t)) \vec{e}_{\mathcal{M}}.
  \label{eq:sol:expl}
\end{equation}

The mass matrix around the minimum is then given by 
\begin{align}
  \mat{m}_{S}^{2} &= \nabla_{\vec{\Phi}} \nabla_{\vec{\Phi}}^{T} V 
 = \mat{M}^{2}_{S} + (4 \pi)^{2} \nabla_{\vec{\Phi}} \mathbb{B} \nabla_{\vec{\Phi}}^{T} t
 \notag
 \\
 &+ (4 \pi)^{2} \nabla_{\vec{\Phi}} t \nabla_{\vec{\Phi}}^{T} \mathbb{B}
 + (4 \pi)^{2} \mathbb{B} \nabla_{\vec{\Phi}} \nabla_{\vec{\Phi}}^{T} t
  \notag 
  \\
  & + \nabla_{\vec{\Phi}} \nabla_{\vec{\Phi}}^{T} \mathbb{A},
\end{align}
where $\mat{M}^{2}_{S}$ is the tree-level scalar mass matrix. Notice that the term proportional to $\mathbb{B}$ is canceled by a similar term resulting from  $\nabla_{\vec{\Phi}} \nabla_{\vec{\Phi}}^{T} \mathbb{A}$.

With the above solution at hand, it is possible to incorporate the corrections previously neglected by expanding $\mathbb{A}$ around the stationary point. This results in finite corrections that shift the values of quartic couplings near the minimum: $\mat{\Lambda}(t) \to \mat{\Lambda}(t) + \Delta \mat{\Lambda}$. It is then possible to use Eq.~\eqref{eq:sol:expl} to compute the corrected minimum solution which, generally, is not exactly aligned with the flat direction indicated by the Gildener-Weinberg approximation.

\section{Detailed Calculation of the Potential Minimum}
\label{app:detail:min}
We now show how the general formalism can be applied, recomputing 
the two-field model used in the main text, where  $\vec{\Phi} = (h, s)$ and the scalar quartic coupling matrix is
\begin{equation}
 \mat{\Lambda}(t) =
  \frac{1}{4}
  \begin{pmatrix}
    \lambda_{H}(t) & \frac{1}{2} \lambda_{HS}(t)
    \\
    \frac{1}{2} \lambda_{HS}(t) & \lambda_{S}(t)
  \end{pmatrix},
\end{equation}
Positivity conditions can be recovered from $\det (\mat{\Lambda})\geq 0$ and from
the adjugate of the scalar quartic matrix
\begin{equation}
 \adj (\mat{\Lambda}) =
  \frac{1}{4}
  \begin{pmatrix}
    \lambda_{S} & -\frac{1}{2} \lambda_{HS}
    \\
    -\frac{1}{2} \lambda_{HS} & \lambda_{H}
  \end{pmatrix}, 
\end{equation}
We choose $\pivot = s$ along the flat direction \ie\
\begin{equation}
  \vec{e}_{\pivot} =   \vec{e}_{s} =   \begin{pmatrix}
    0 \\ 1
  \end{pmatrix},
\end{equation}
Then  $t = \ln (s^{2}/s_{0}^{2})/(4 \pi)^{2}$. 
We identify $s_{0}$ with the field value at which the potential crosses the critical boundary $V = 0$ of the full potential, \ie, the scale corresponding to the tree-level flat direction.
Around the multi-phase point we can neglect wave-function renormalisations, \ie\
the $\nabla_{\vec{\Phi}} \mathbb{A}$ in Eq.~\eqref{eq:minmf}.
The minimum scale is given by the usual Gildener-Weinberg relation, $s^{2} = e^{-\frac{1}{2}} s_{0}^{2}$.    

In order to account for the one-loop corrections,
we can expand $\mathbb{A}$ in the parameter $h^2/s^2$, small in a region of the field space close to the identified flat direction. By defining $t_S = t(s=s_S)$ and $t_{HS}=t (s=s_{HS})$, the result of the expansion can be incorporated in the definition of the field-dependent couplings, yielding 
\begin{align}
  \lambda_{S}(t) &= \beta_{S} (t - t_{S}),
  \\
  \lambda_{HS}(t) &= \beta_{HS} (t - t_{HS}),
\end{align}
which recovers Eq.~\eqref{eq:run:lambda} in the main text.

Then, the parameter $t_{0}$ corresponding to the scale $s_{0}$ can be obtained from the flat direction equation
\begin{equation}
\begin{split}
  0 &= \det (\mat{\Lambda} (t_{0}) )
  \\
  & = \beta_{HS}^{2} t_{0}^{2} - (4 \lambda_{H} \beta_{S} + 2 \beta_{HS}^{2} t_{HS}) t_{0} + 4 \lambda_{H} \beta_{S} t_{S}.
 \end{split}
\end{equation}
Ignoring higher order terms proportional to $\beta_{HS}^{2}$ (irrelevant unless $\abs{t_{HS} - t_{S}}$ is  large), 
the quadratic equation reduces to $\lambda_{S}(t_{0}) \approx 0$ and we find that $s_{0} \approx s_{S}$.

The Hadamard square of the vector indicating the corrected location of the potential minimum in field space is then computed as 
\begin{equation}
  \vec{\Phi}^{\circ 2} 
  =
  e^{-\frac{1}{2}} s_{0}^{2} \frac{\adj (\mat{\Lambda}(t)) \vec{e}_{s}}{\vec{e}_{s}^{T} \! \adj (\mat{\Lambda}(t)) \vec{e}_{s}}.
\end{equation}
From the results of \ref{app:matrix:min}, we have that $t = t_{0} -(1/2) / (4 \pi)^{2} \approx t_{S} -(1/2) / (4 \pi)^{2}$ at the minimum, and $\ln R \approx (4 \pi)^{2} (t - t_{HS})$ \ie\ $R = e^{-1/2} s_S^2/s_{HS}^2$. Using this to express $t_{HS}$ in terms of $\ln R$, the potential minimum is at
\begin{equation}
  \vec{\Phi}^{\circ 2} 
  \approx
  e^{-\frac{1}{2}} s_{S}^{2}
    \begin{pmatrix}
    - {\beta_{\lambda_{HS}} \ln R}/{2 \lambda_{H} (4 \pi)^{2}}
     \\ 
     1
  \end{pmatrix}.
\end{equation}
From the scalar mass matrix at the minimum, we find that the mixing angle is given by
\begin{equation}
\begin{split}
  \theta &\approx \frac{(\mat{m}_{S}^{2})_{hs}}{m_{S}^{2} - m_{h}^{2}} \approx \frac{\beta_{HS} (1 + \ln R)}{2 \beta_{S} + \beta_{HS} \ln R} \frac{h}{s}.
\end{split}
\end{equation}
As this mixing is small, the mass eigenvalues are approximatively given by the diagonal elements
of the mass matrix.
It is then easy to see that since $\nabla_{\vec{\Phi}} t \propto \vec{e}_{s}$, for the Higgs mass only the tree-level contribution with running couplings (amended by corrections from $\mathbb{A}$) does not vanish:
\begin{equation}
\begin{split}
  m_{h}^{2} &= 3 \lambda_{H}(t) h^{2} + \frac{1}{2} \lambda_{HS}(t) s^{2} 
  \\
  &\approx \frac{-s^2 \beta_{\lambda_{HS}}\ln R}{(4\pi)^2} = 2\lambda_H h^2.
  \end{split}
\end{equation}
Since $\vec{\Phi}^{T} (\mat{M}_{S}^{2} + \nabla_{\vec{\Phi}} \nabla_{\vec{\Phi}}^{T} \mathbb{A}) \vec{\Phi} = 12 V$, the scalon mass is 
\begin{equation}
  m_{s}^{2} \approx \frac{1}{s^{2}} \vec{\Phi}^{T} \mat{m}^{2} \vec{\Phi} \approx \frac{\beta_{\lambda_S}}{(4\pi)^2} 2s^2.
\end{equation}
We have thereby recovered the expressions obtained in the main text from direct computations.

\bibliographystyle{elsarticle-num}
\bibliography{artCW}

\end{document}